\documentclass{appolb}
\usepackage{graphicx}

\begin{document}
\pagestyle{plain}
\newcount\eLiNe\eLiNe=\inputlineno\advance\eLiNe by -1
\title{ONCE AND TWICE SUBTRACTED DISPERSION RELATIONS IN THE ANALYSIS OF  $\pi\pi$ AMPLITUDES
}
\author{R. KAMI\'NSKI\footnote{Speaker}$^a$, R. GARCIA-MARTIN$^{b}$, J. R. PELAEZ$^c$ and \\ F. J. YNDURAIN$^d$
\address{$^a$Institute of Nuclear Physics PAN, Krak\'ow, Poland\\
$^b$Institut de Physique Nucl\'eaire (IN2P3-CNRS), Orsay, France.\\
$^c$Departamento de F\'{\i}sica Te\'orica II, Universidad Complutense de Madrid, Spain\\
$^d$Departamento de F\'{\i}sica Te\'orica, Universidad Aut\'onoma de Madrid, Spain
}}

\maketitle

\begin{abstract}
Once and twice subtracted crossing symmetric dispersion relations
applied to $\pi\pi \to \pi\pi$ scattering data are analyzed and compared.
Both sets of dispersion relations can 
be used to test the $\pi\pi$ amplitudes in low partial waves up to about 1 GeV.
We show how once subtracted 
dispersion relations can provide stronger 
constraints for $\pi\pi$ amplitudes than 
twice subtracted ones in the 400 to 1100 MeV range, 
given the same experimental input.
\end{abstract}

\section{Introduction}
A set of dispersion relations for $\pi\pi \to \pi\pi$ 
scattering amplitudes incorporating crossing symmetry 
were presented by Roy in 1971 \cite{Roy71} (hereafter called the Roy equations).
Two subtractions were used for faster convergence of the dispersive
integrals. These equations are a very efficient tool 
in testing the $\pi\pi$ experimental data.

In the last few years, several analysis of $\pi\pi$ amplitudes using
Roy equations have appeared with different aims. For instance, to
predict the low energy $S$ and $P$ waves below 800 MeV with the aid of
Chiral Perturbation Theory 
(ChPT), which is used to fix the threshold behavior of amplitudes \cite{A4,CGLNPB01}.
Or, for example, to eliminate the long standing "up-down" 
ambiguity in scalar-isoscalar $\pi\pi$ amplitudes below 1 GeV \cite{klr}.
Later on, they were used by our group, together with forward dispersion relations (FDR)
to describe data and also test predictions of ChPT \cite{KPYII,KPYIII}.
All these works provide very precise 
determinations of $\pi\pi$ amplitudes for the $S$ and $P$ waves
below about 1 GeV and very precise predictions or determinations of scattering 
lengths and parameters of the $\sigma$ meson \cite{KPYIII,Caprini,GPY}.

More recently,  together with our FDR and Roy 
equations analyses \cite{KPYIII} we have developed 
a set of once subtracted 
dispersion relations that also incorporate $\pi\pi$ crossing symmetry  
(hereafter called GKPY). 
Due to the single subtraction, the higher $\pi\pi$ partial waves 
and the high energy region contributions are less 
suppressed in these equations than in Roy's, although
 integrals in the GKPY equations still converge. 

In this presentation we briefly analyze and 
compare the structure and cancellations 
that occur in Roy equations versus those in the GKPY eqs.

\section{Twice subtracted dispersion relations (Roy equations).}

Roy equations can be expressed as a sum
\begin{equation}
{\rm Re} f_l^I(s) = ST(s) + KT(s) + DT(s),
\label{Decomposition}
\end{equation} 
where $ST(s)$, $KT(s)$ and $DT(s)$ are called subtraction, kernel and driving 
terms, respectively.
The ${\rm Re} f_l^I(s)$ on the left hand side is called ``output'' amplitude
throughout this paper
and can be calculated for the $S$ and $P$ waves up to about 1 GeV.
The subtraction terms read
\begin{equation}
ST(s)=  a_{0}^{0}\delta_{I0} \delta_{\ell 0} + a_{0}^{2}\delta_{I2} \delta_{\ell 0} +
        \displaystyle \frac{s-4m_{\pi}^2}{12}(2a_{0}^{0}-5a_{0}^{2})
       (\delta_{I0}\delta_{\ell 0}+
        \frac{1}{6}\delta_{I1}\delta_{\ell 1} 
        -\ \frac{1}{2}\ \delta_{I2}\delta_{\ell 0})
\label{ST.Roy}
\end{equation}
where $a_{0}^{0}$ and $a_{0}^{2}$ are the  $S0$ and $S2$
scattering lengths. Let us remark that
the $ST(s)$ contain a piece that grows with $s-4m_\pi^2$.
The $KT(s)$ are singular integrals of some known kernels $K_{\ell \ell^\prime}^{I I^\prime}(s,s')$  multiplied by
imaginary parts of the ``input'' amplitudes ${\rm Im }f_{\ell'}^{I^\prime}(s')$
\begin{equation}
KT(s) = \displaystyle \sum\limits_{I'=0}^{2}
        \displaystyle \sum\limits_{\ell '=0}^{1}
     \hspace{0.25cm}-\hspace{-0.55cm}
        \displaystyle \int \limits_{4m_\pi^2}^{s_{max}}\ ds'
     K_{\ell \ell^\prime}^{I I^\prime}(s,s') \, \mbox{Im }f_{\ell'}^{I^\prime}(s').
\label{KT.Roy}
\end{equation}
For twice subtracted dispersion relations the kernels are proportional to 
$1/s'^2(s'-s)$. 
In our analysis 
(see for example \cite{KPYIII}) $s_{max}^{1/2} = 1.42$ 
GeV, which is the maximum energy
for which we can 
parameterize experimental $S$ and $P$ wave 
data in terms of phase shifts and inelasticities.
The so called driving terms $DT(s)$ collect
the $s^{1/2} > 1.42$ GeV contributions 
from all waves -- parameterized in terms of
Regge theory -- together with
the contributions from $\ell\geq 2$ partial waves below 1.42 GeV.
Our fits of $\pi\pi$ amplitudes to 
data are then constrained to minimize, within uncertainties,
the difference between 
``input'' and ``output'' amplitudes for the $S$,  and $P$ waves at 25
equidistant energy values $s_i^{1/2}$ below 1 GeV~\cite{KPYIII}.

A similar method using only forward dispersion relations
was used in~\cite{PY} to show that 
many of the presently available experimental data sets dot not fulfill 
well enough the analyticity constraints.

\section{Once subtracted dispersion relations (GKPY equations)}

The general structure of the GKPY equations is similar to that of the Roy 
equations in Eq. (\ref{Decomposition}).
However, the 
subtraction terms are now constant for each wave 
and are expressed as combinations of scattering lengths
\begin{equation}
ST(s) =  \sum_{I'}C^{st}_{II'}a_0^{I'} 
\label{ST.GKPY} 
\vspace*{-0.3cm}
\end{equation}
with
$a_0 = (a_0^0, 0, a_0^2)$ and $C^{st}$ the usual crossing matrix. 
Since there is just one subtraction, the integral kernels
are now proportional to $1/s'(s'-s)$.  Consequently, 
the high energy contributions to kernel and particularly to the
driving terms are now 
less suppressed compared with the Roy equations case.
Nevertheless, we will see in the next section 
that the effect of driving terms is still 
smaller than the kernel terms for the $P$, $S2$ and $S0$ 
waves in almost the whole energy region of interest, 
which allows us to obtain a reliable calculation.

\section{Comparison of the Roy and GKPY dispersion relations}

In Table (\ref{Table.Thr.Exp}) we compare 
the structure of the different terms of Roy and GKPY equations
for threshold parameters (TP), defined as:
\begin{equation}
  Re f_{\ell}^I(s\approx 4m_{\pi}^2) = (s-4m_{\pi}^2)^{\ell} \left[a_{\ell}^I + b_{\ell}^I(s-4m_\pi^2) + ...\right].
\label{Thr.Exp} 
\end{equation}

\begin{table}[b!]
\begin{center}
\begin{tabular}[b]{|c|c|r|c|c|c|}  
\hline
& & \multicolumn{2}{|c|}{Roy} & \multicolumn{2}{|c|}{GKPY} \\
wave & TP & \multicolumn{1}{|c|}{$ST$} & $KT \& DT$ & $ST$ & $KT \& DT$ \\ 
\hline
&&&&& \\
$S0$ &$a_0^0$& $a_0^0 + \alpha(s-4)$ & $\beta(s-4)$ &
$a_0^0 + 5a_0^2$ & $\delta(s-4)-5a_0^2 $\\
&&&&& \vspace{-0.3cm}\\
$P$&
 $0$ & \hspace{0.8cm}$\alpha'(s-4)$ & $\beta'(s-4)$  & 
$a_0^0 - \frac{5}{2}a_0^2$ &$\delta'(s-4)-a_0^0 + \frac{5}{2}a_0^2$\\
&&&&& \vspace{-0.3cm}\\
$S2$&
 $a_0^2$ & $a_0^2 +\alpha''(s-4)$ & $\beta''(s-4)$ & 
$a_0^0 + \frac{1}{2}a_0^2$ & $\delta''(s-4)-a_0^0+\frac{1}{2}a_0^2 $ \\
&&&&& \\
\hline
\end{tabular}
\end{center}
\vspace{-0.5cm}
\caption{Comparison of the threshold expansion
 parameters (TP) of the "output" amplitudes for Roy and
GKPY equations ($m_\pi$ units). Greek letters stand for constants
whose precise value is irrelevant for the discussion. Note that 
the GKPY value at threshold is obtained from a 
cancellation between $ST,KT$ and $DT$ terms.
}
\label{Table.Thr.Exp} 
\end{table}
As  is seen from the Table (\ref{Table.Thr.Exp}) 
in the case of Roy equations, not only the linear terms in $ST(s)$,
but also the whole kernel and driving terms vanish at
$s=4m_{\pi}^2$.
Thus, the constant terms in $ST(s)$ are the ones that ensure
 the correct values of the threshold parameters in the "output" amplitudes.
In contrast, for the GKPY equations 
the $KT(s)$ and $DT(s)$ do not vanish at 
threshold and it is a combination of their nonzero constant parts 
with the $ST(s)$ terms that yield the correct threshold value.
Hence, GKPY uncertainties around threshold are expected to be larger
than for Roy equations.

\begin{figure}[ht!]
\vspace{9pt}
\includegraphics*[width=6.0cm]{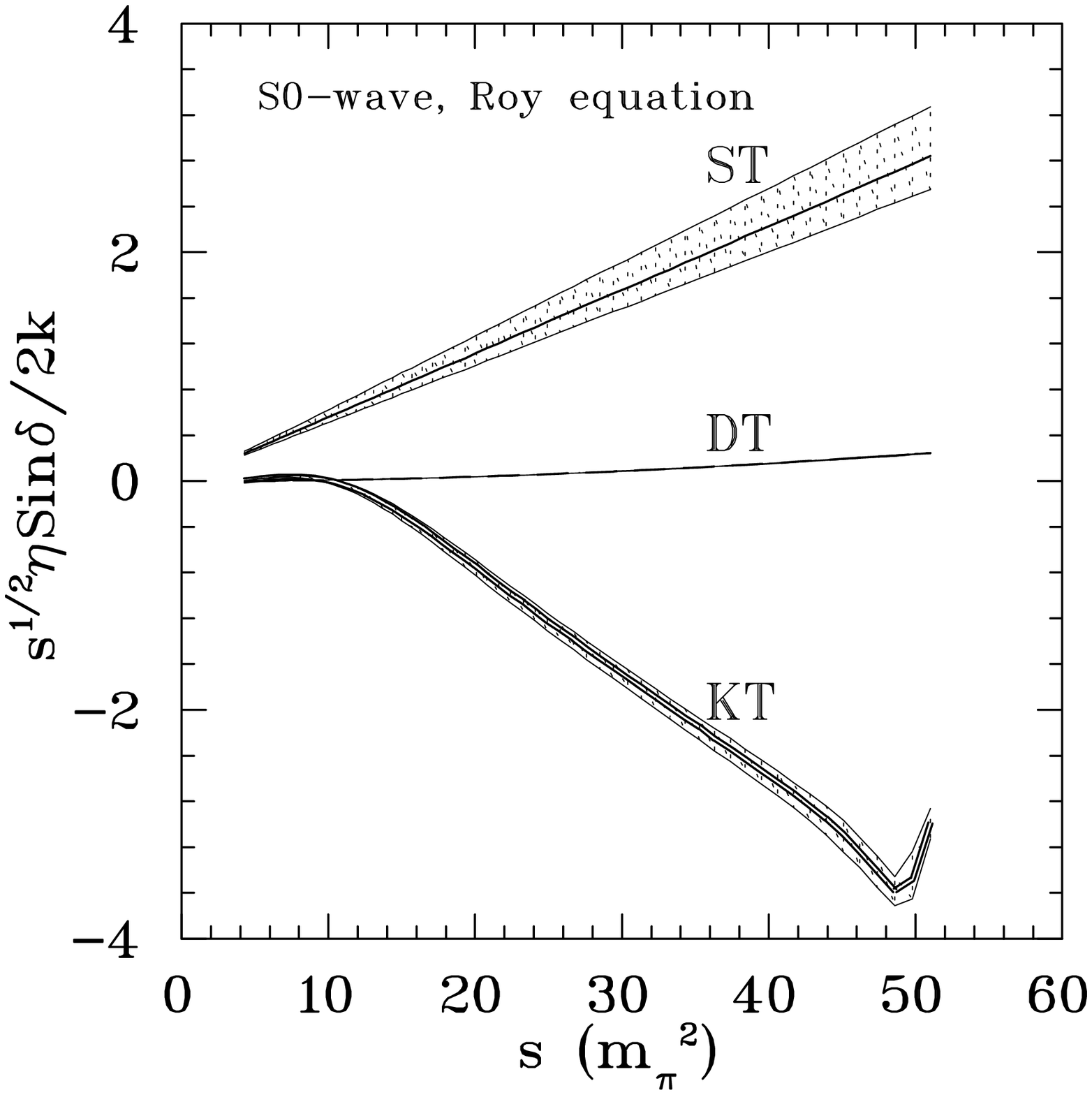}
\includegraphics*[width=6.0cm]{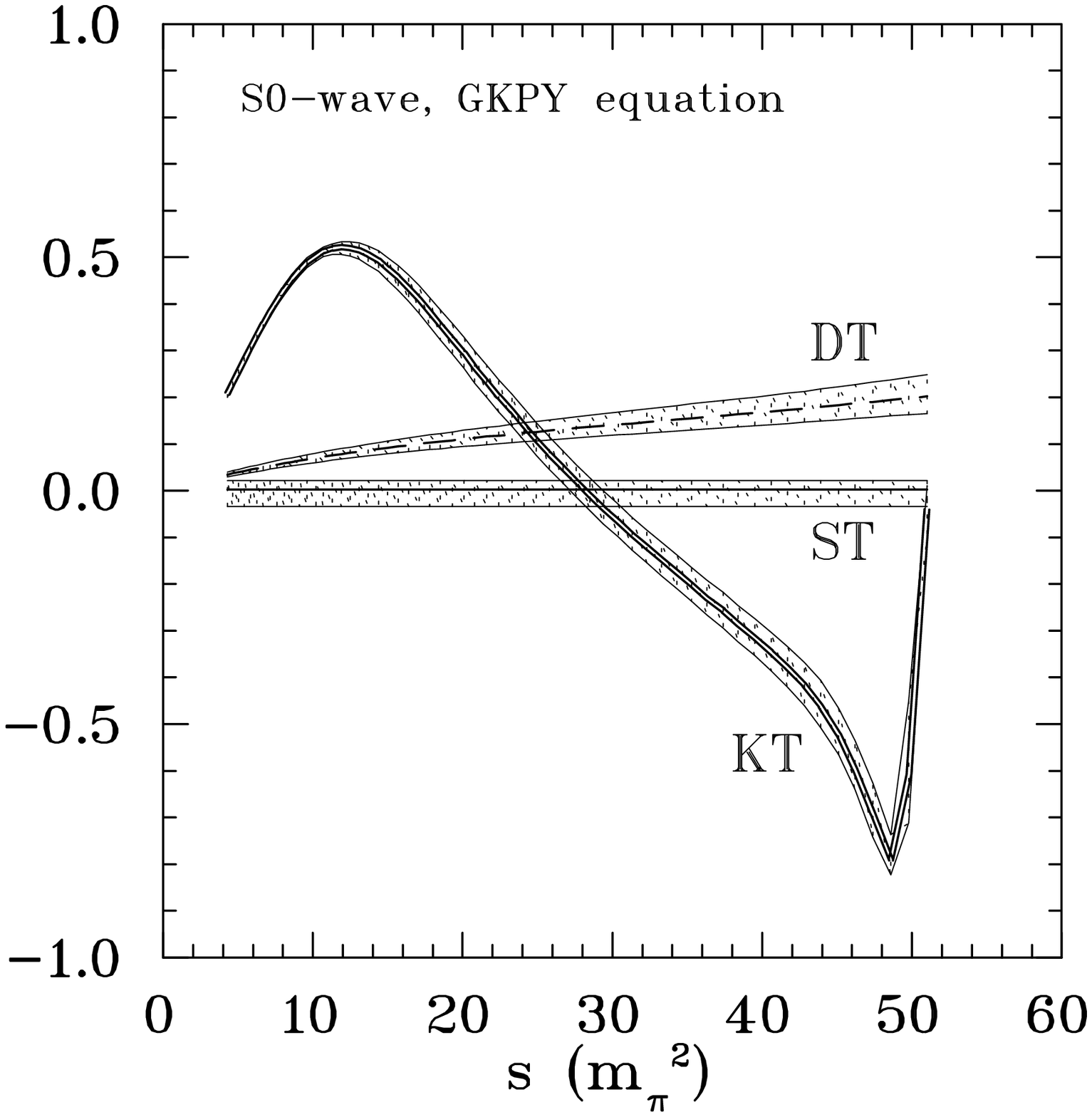}
\includegraphics*[width=6.0cm]{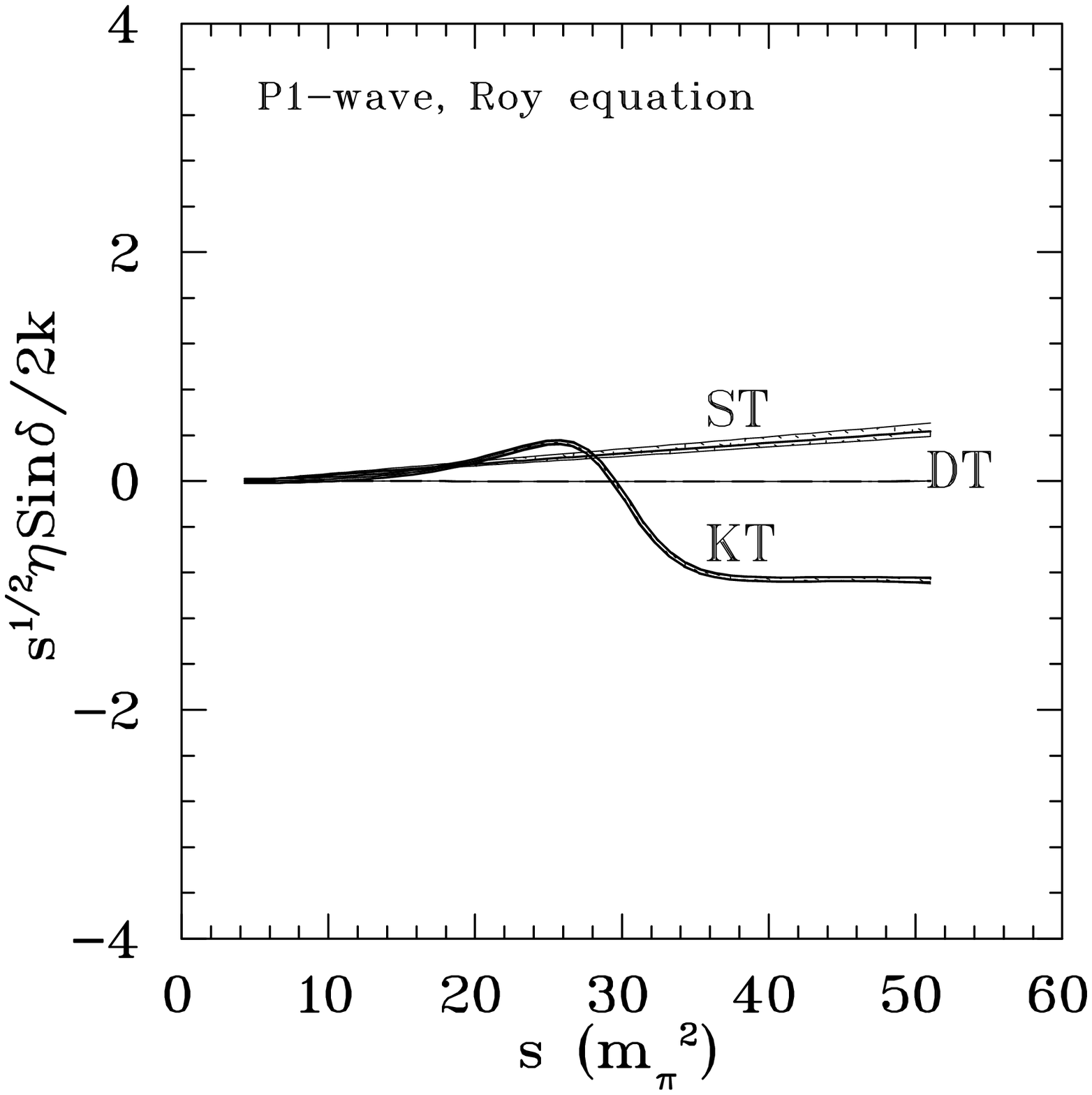}
\includegraphics*[width=6.0cm]{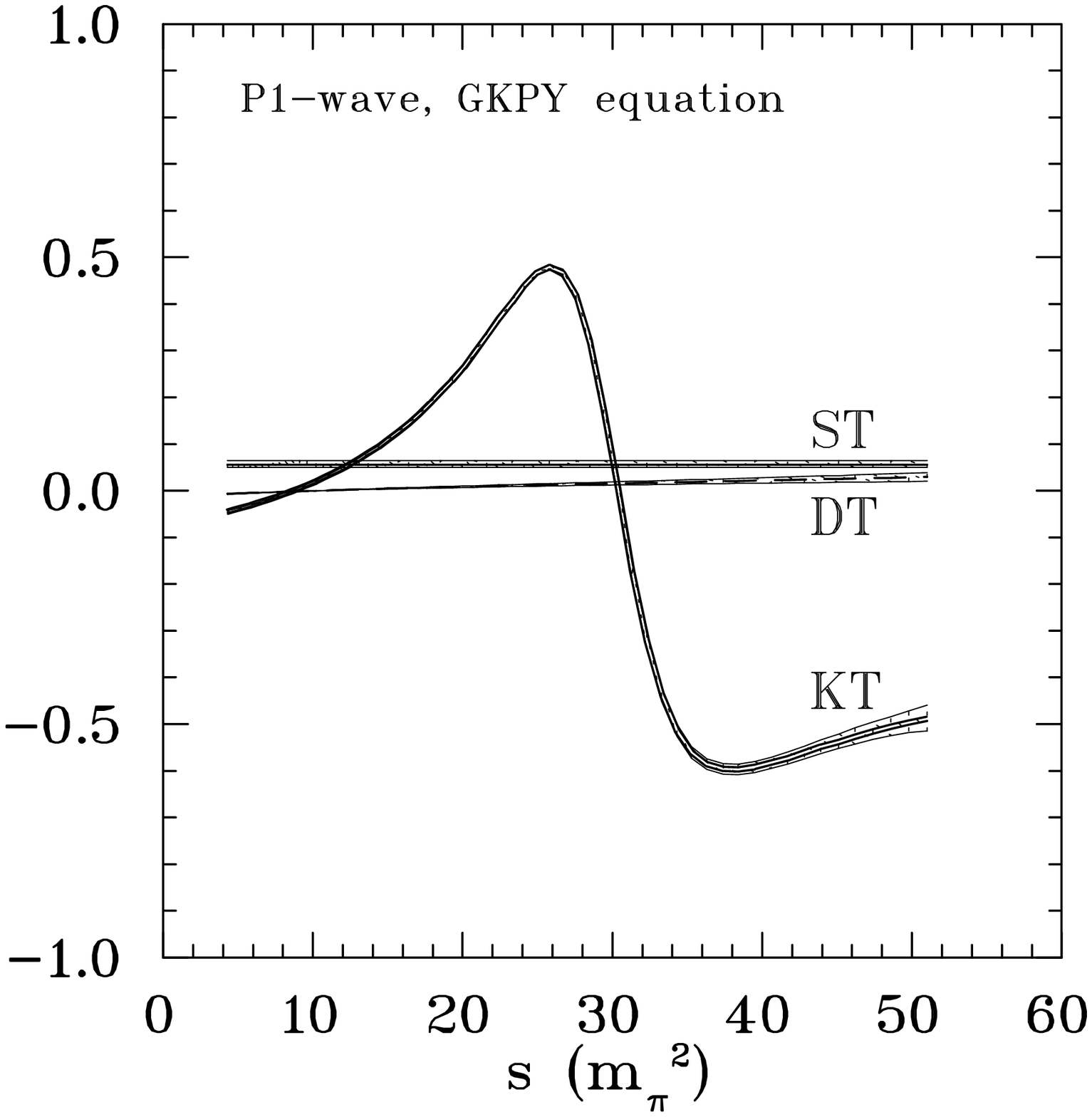}
\includegraphics*[width=6.0cm]{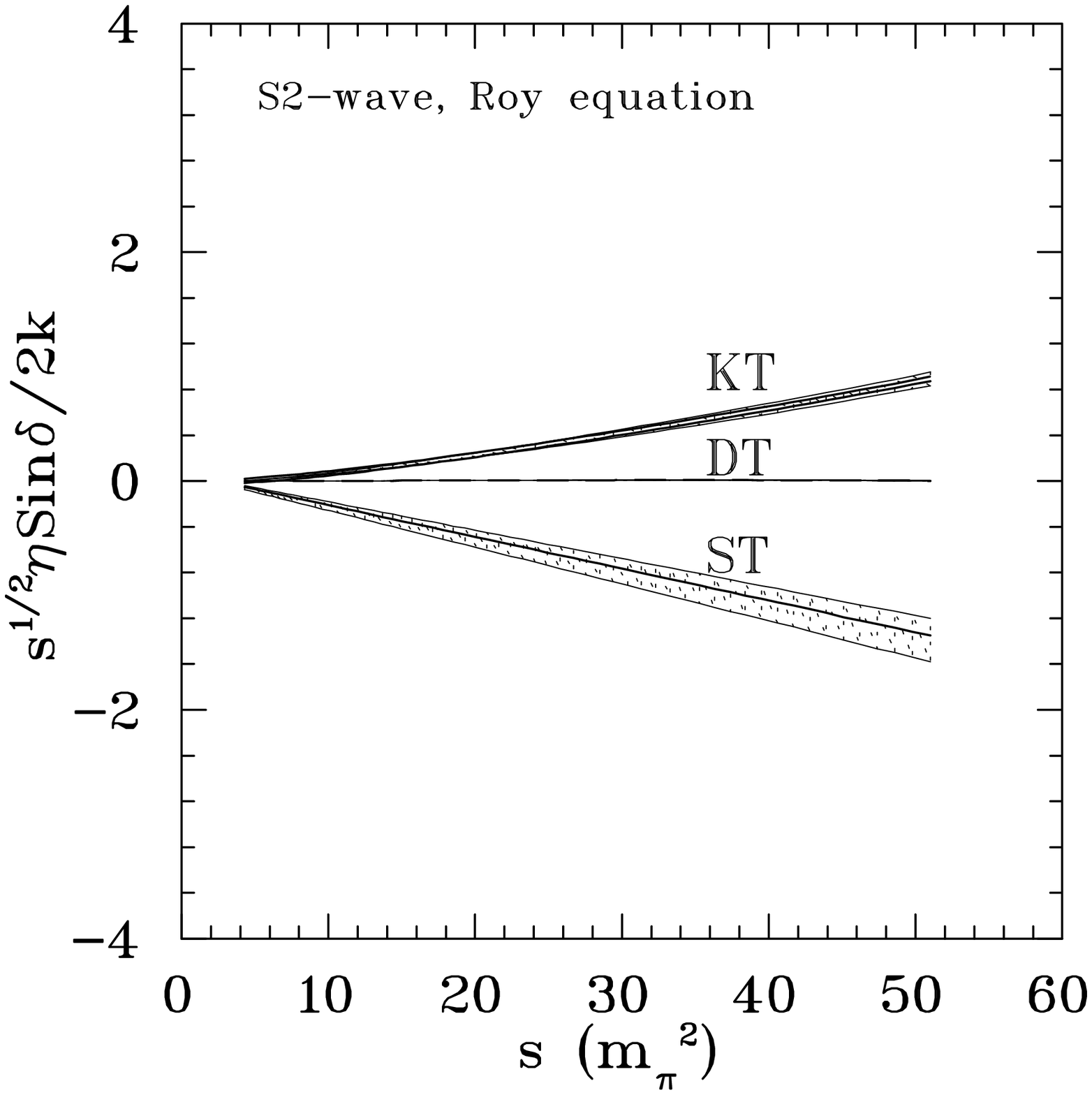}
\hspace{0.5cm}\includegraphics*[width=6.0cm]{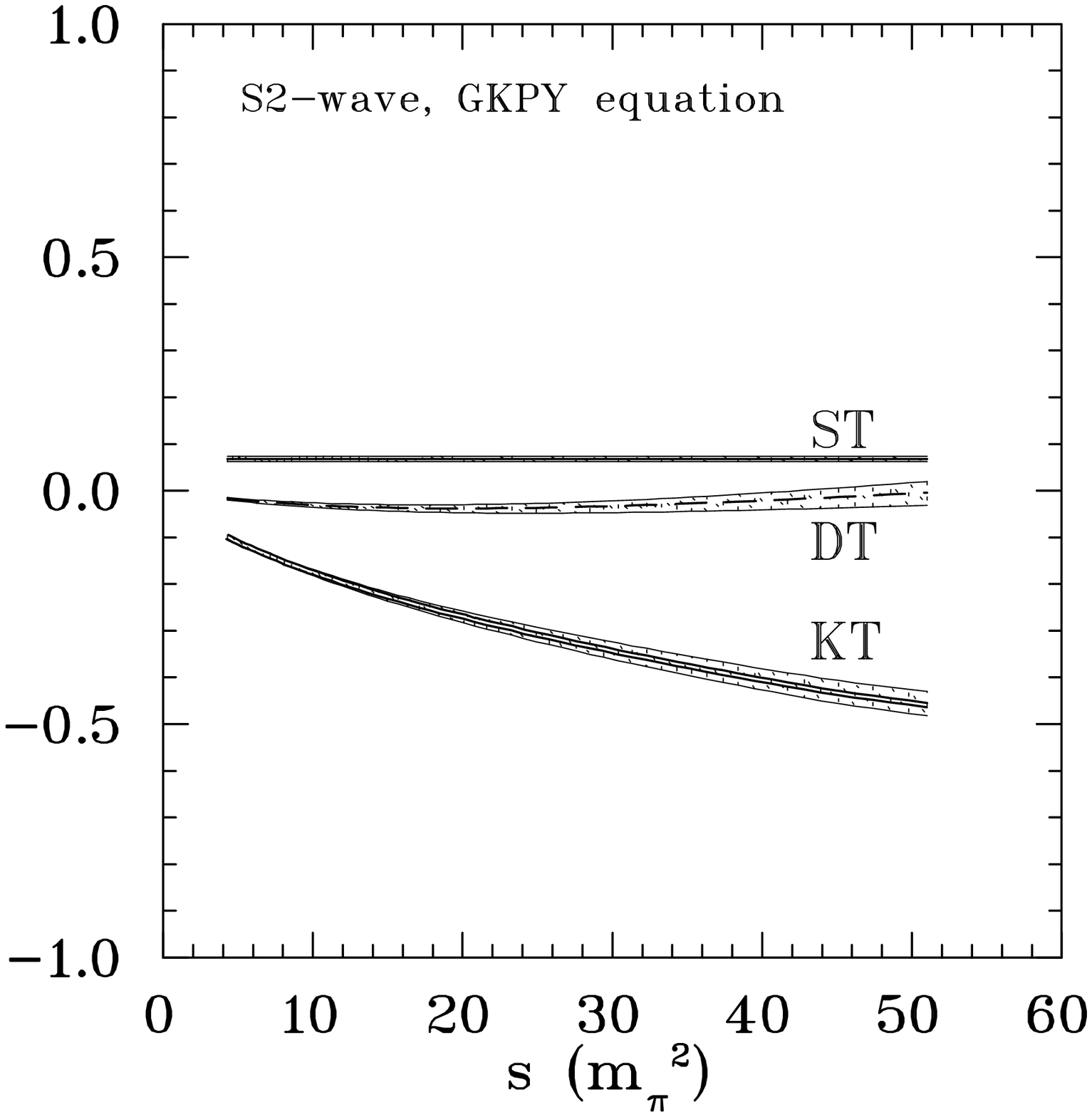}
\caption{Subtraction ($ST$), kernel ($KT$) and driving ($DT$) terms for  
the $S0$, $P$ and $S2$ waves from Roy equations (left panel) and
from the GKPY ones (right panel).
Dashed bands denote the errors of these terms. Note that we use $m_\pi$ units.}
\label{fig:decomposition}
\end{figure}

\begin{figure}[ht!]
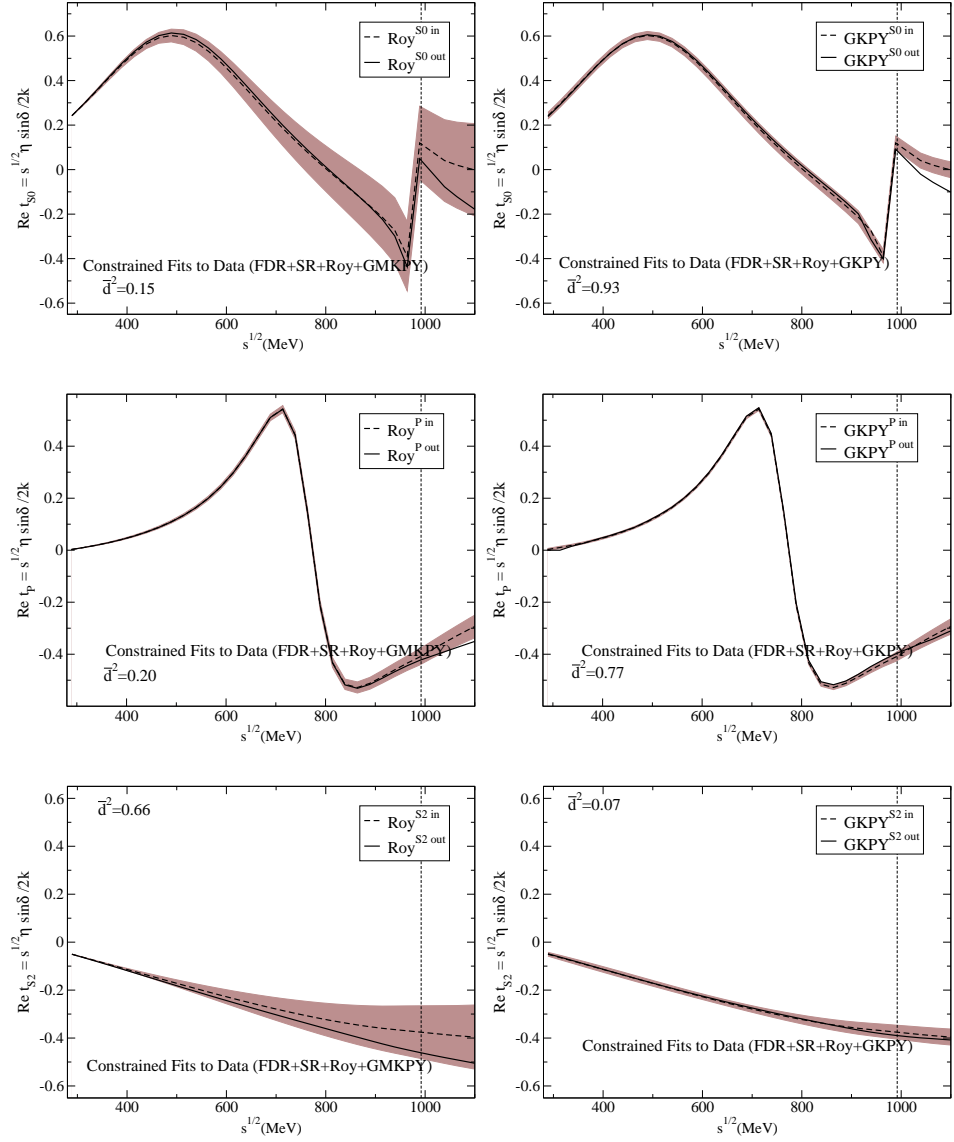

\vspace{9pt}
\includegraphics*[width=6.2cm]{./figuresfromRuben/cdf-roys0all.eps}
\includegraphics*[width=6.2cm]{./figuresfromRuben/cdf-pacos0all.eps}

\vspace{0.5cm}

\includegraphics*[width=6.2cm]{./figuresfromRuben/cdf-roypall.eps}
\includegraphics*[width=6.2cm]{./figuresfromRuben/cdf-pacopall.eps}

\vspace{0.5cm}

\includegraphics*[width=6.2cm]{./figuresfromRuben/cdf-roys2all.eps}
\includegraphics*[width=6.2cm]{./figuresfromRuben/cdf-pacos2all.eps}
\caption{Comparison of preliminary results from 
a data fit constrained to satisfy Forward dispersion relations, 
Roy and GKPY equations for the $S0$, $P$ and $S2$ waves. 
The dark bands show errors of these equations. 
Dashed and solid lines represent "input" and "output" 
amplitudes respectively. The $\bar d^2$ are averaged values of 
$\chi^2$ for the Roy or GKPY equation corresponding to given wave.}
\label{fig:Results}
\end{figure}

In Fig.~(\ref{fig:decomposition}) 
we show the decomposition of Roy and GKPY equations 
into subtraction, kernel and
driving terms.
Error bands are obtained from a Monte Carlo
with $10^5$ Gaussian samplings, 
within three standard deviations of their central values,
of all parameters used to describe 
the different input (for details see \cite{KPYIII}). 
Let us remark that, above 
$\sqrt{s}\simeq 450\,$MeV ($s\simeq10$ in $m_\pi$ units),
 the $ST(s)$ and $KT(s)$ terms
in Roy equations suffer a very strong cancellation to give the total
output amplitudes that, as seen in Fig.(\ref{fig:Results}), satisfy
 $\mbox{Re } f^I_{\ell} < 0.6$ (note that the vertical
 scale for the Roy equations figures is much bigger than 1). 
In contrast, for GKPY equations, the kernel terms are 
dominant and there is no such strong cancellation. It is also clear from Fig.(\ref{fig:decomposition}) 
that, above $\sqrt{s}\simeq400\,$ MeV, 
the errors in Roy equations are significantly bigger
than those in the GKPY ones. 
The main source of Roy equations uncertainties 
is the experimental error on $a_0^{2}$, that propagates 
through a term in $ST(s)$ proportional to $s$.
In contrast,
the $ST(s)$ and their errors in the GKPY equations are constant.

Fig.(\ref{fig:Results}) compares the size of the uncertainties for
a preliminary data fit constrained to satisfy 
forward dispersion relations,
 sum rules, 
Roy and GKPY equations (see \cite{KPYIII}).
We also give preliminary values of an average $\chi^2/NP$ 
for each wave, where $NP = 29$ is number of points where "input" and
"output" amplitudes are compared.

In conclusion, for the same input, GKPY equations 
have much smaller errors than standard Roy equations 
above $\sqrt s \approx $ 400 MeV and 
thus they become a
very promising tool to obtain a precise
data analysis of  $\pi\pi$ 
amplitudes in the 400 to 1100 MeV region.

\end{document}